\documentclass[12pt]{article}
\usepackage{tikz}
\usetikzlibrary{arrows,decorations.markings,mindmap,patterns}

\newlength{\oldoddsidemargin}
\oldoddsidemargin=\oddsidemargin \oddsidemargin=\evensidemargin
\evensidemargin=\oldoddsidemargin

\makeatletter
\def\cleardoublepage{\clearpage\if@twoside \ifodd\c@page\else
   \hbox{}
   \thispagestyle{empty}
   \newpage
   \if@twocolumn\hbox{}\newpage\fi\fi\fi}
\makeatother \clearpage{\pagestyle{plain}\cleardoublepage} 
\usepackage{tabularx}
\usepackage{longtable}
\usepackage{multirow}
\usepackage{graphicx, color}
\usepackage{subfigure}
\usepackage{amsmath}
\usepackage{amsfonts}
\usepackage{amssymb}
\usepackage[OT2,T1]{fontenc}
\usepackage[OT1]{fontenc}
\DeclareSymbolFont{cyrletters}{OT2}{wncyr}{m}{n}
\DeclareMathSymbol{\Lo}{\mathalpha}{cyrletters}{"4C}
\usepackage[english,german]{babel}
\usepackage{bbding}

\usepackage{float}
\usepackage{dsfont}   
\usepackage{hyperref}
\usepackage{mathrsfs} 
\usepackage{makeidx}
\usepackage{listings}
\makeindex
\usepackage{setspace}
\newcolumntype{C}[1]{>{\centering\arraybackslash}b{#1}}
\newcolumntype{D}[1]{>{\centering\arraybackslash}b{#1}}
\newcolumntype{R}[1]{>{\raggedleft\arraybackslash}b{#1}}                                                                                                                                                                                                                                                                                                                                                                                                                                                                                                                                                                                                                                                                                                                                                                                                                                                                                                                                                                                                                                                                                                                                                                                                                                                                                                                                                                                                                                                                                                                                                                                                                                                                                                                                                                                                                                                                                                                                                                                                                                                                                                                                                                                                                                                                                                                                                                                                                                                                                                                                                                                                                                                                                                                                                                                                                                                                                                                                                                                                                                                                                                                                                                                                                                                                                                                                                                                                                                                                                                                                                                                    
\newcolumntype{M}[1]{>{\centering\arraybackslash}m{#1}}

\def\d{\mathrm d}

\def\bea{\begin{eqnarray}}
\def\eea{\end{eqnarray}}

\def\I{\mathrm{i}}

\def\bsm{\left( \!\begin{smallmatrix}}
\def\esm{\end{smallmatrix} \!\right)}

\newcommand{\mac}[1]{\mathcal{#1}}

\newcommand{\nn}{\nonumber}

\newcommand\cF{{\mac{F}}}

\numberwithin{equation}{section}

\renewcommand{\a}{\alpha}

\renewcommand{\th}{\theta}
\newcommand{\Th}{\Theta}

\newcommand{\pt}{\partial}

\newcommand{\be}{\begin{equation}}
\newcommand{\ee}{\end{equation}}
\newcommand{\ba}{\begin{align}}
\newcommand{\ea}{\end{align}}

\def\cR{\mac{R}}
\def\cD{\mac{D}}

\def\cF{\mac{F}}

\def\bcD{\bar{\mac{D}}}
\def\Phid{\Phi^\dagger}

\begin{document}
\allowdisplaybreaks
\selectlanguage{english}

\begin{titlepage}

\title{Higher-Derivative Chiral Superfield Actions Coupled to $\mac{N}=1$ Supergravity {\LARGE \\[.5cm]  }}
\author{{Michael Koehn${}^{1}$, Jean-Luc Lehners${}^{1}$ and Burt A. Ovrut${}^{2}$}\\[5mm]
{\it ${}^{1}$ Max-Planck-Institute for Gravitational Physics} \\
{\it Albert Einstein Institute, 14476 Golm, Germany}\\[4mm]
{\it  ${}^{2}$ Department of Physics, University of Pennsylvania} \\
   {\it Philadelphia, PA 19104--6396}}
\date{}
\maketitle

\begin{abstract} 

\let\thefootnote\relax\footnotetext{michael.koehn@aei.mpg.de,~~~jlehners@aei.mpg.de,~~~ovrut@elcapitan.hep.upenn.edu} We construct $\mac{N}=1$ supergravity extensions of scalar field theories with higher-derivative kinetic terms. Special attention is paid to the auxiliary fields, whose elimination leads not only to corrections to the kinetic terms, but to new expressions for the potential energy as well. For example, a potential energy can be generated even in the absence of a superpotential. Our formalism allows one to write a supergravity extension of any higher-derivative scalar field theory and, therefore,  has applications to both particle physics and cosmological model building. As an illustration, we couple the higher-derivative DBI action describing a 3-brane in 6-dimensions to $\mac{N}=1$ supergravity. This displays a number of new features-- including the fact that, in the regime where the higher-derivative kinetic terms become important, 
the potential tends to be everywhere negative.

\vspace{.3in}
\noindent
\end{abstract}

\thispagestyle{empty}

\end{titlepage}

\tableofcontents

\section{Introduction}

Since its discovery \cite{Golfand:1971iw,Volkov:1973ix,Wess:1974tw}, supersymmetry has been investigated with enthusiasm by theoretical physicists. Representations of the supersymmetry algebra contain bosonic and fermionic degrees of freedom in equal numbers. Moreover, particles belonging to the same representation have equal mass. Since superpartners with the same mass as conventional particles have not been observed, four-dimensional supersymmetry cannot be an unbroken low energy symmetry. Nevertheless, there are good reasons to take seriously the idea that supersymmetry-- particularly four-dimensional ${\cal{N}}=1$ supersymmetry --might be relevant at higher energies. For example, when ${\cal{N}}=1$ supersymmetry is taken into account, the gauge couplings of the electroweak and strong forces unite to good precision at high energies \cite{Langacker:1983cj}, suggesting the existence of supersymmetric grand unification. Moreover, supersymmetric theories enjoy special finiteness properties that help to explain the hierarchy between the electroweak and the unification/gravitational scales \cite{Dimopoulos:1981zb,Dimopoulos:1981yj}. Last, but not least, ${\cal{N}}=1$ supersymmetry is a central feature of phenomenologically realistic string theories--see, for example \cite{Braun:2005nv,Lukas:1998yy}. 

All of this motivates studying early universe cosmology within the context of ${\cal{N}}=1$ supersymmetry. Since cosmology quintessentially involves gravitation, such theories must be constructed using ``local'' supersymmetry-- that is, ${\cal{N}}=1$ supergravity --and not the ``global'' supersymmetry of low energy particle physics models. This has been done within the context of two-derivative kinetic theories, both in local quantum field theory and superstrings. More recently, however, it has become clear that higher-derivative theories of cosmology are potentially important. These include so-called DBI inflation \cite{Silverstein:2003hf}, ekpyrotic theories with brane collisions \cite{Khoury:2001wf,Khoury:2001bz} and ghost-condensation \cite{Buchbinder:2007ad,Buchbinder:2007tw,Buchbinder:2007at}, as well as other cosmologies constructed on the worldvolume of three-branes \cite{Khoury:2012dn,Ovrut:2012wn}. Motivated by this, in this paper we will develop a framework for constructing higher-derivative kinetic theories of chiral superfields coupled to ${\cal{N}}=1$ supergravity. As a first application of this formalism, we present an example of supergravitational DBI inflation.

This paper builds on previous work \cite{Khoury:2010gb,Khoury:2011da} on globally supersymmetric higher-derivative scalar field theories-- extending it to local ${\cal{N}}=1$ supergravity. We first construct a supergravity version of $(\pt\phi)^4$, the square of the usual kinetic energy of a real scalar field. In the present work, we neglect fermions because a) they are typically unimportant in models of early universe cosmology and b) since their inclusion greatly complicates all equations.  Instead, we focus on the physics of the scalar bosons and the associated auxiliary fields. We will present the fermionic terms, and discuss their role, in forthcoming publications \cite{forthcoming}. When the fermions are set to zero, our supergravity extension of $(\pt\phi)^4$ has a special-- perhaps unique --property; namely, it can be multiplied by an arbitrary function of the scalar fields and their spacetime derivatives, while not altering the pure supergravity sector of the Lagrangian. Because this multiplicative factor is arbitrary, our formalism allows one to write a supergravity extension of {\it any} higher-derivative Lagrangian built out of scalar fields and their spacetime derivatives. 

As always in supergravity, a special role is played by the auxiliary fields. In this paper, we devote considerable attention to their properties. In ordinary two-derivative chiral supergravity, elimination of the auxiliary fields leads to a well-known formula for the potential $V$. In terms of a K\"{a}hler potential $K$ and superpotential $W$ \cite{Cremmer:1978hn}, this is given by
\be
V = e^K \left( K^{,A^{i}A^{j*}} |D_{A^{i}} W|^2 - 3 |W|^2 \right) \ ,
\ee
where $A^{i}$ denotes the complex scalar component of a chiral supermultiplet. In higher-derivative supergravity theories, we find two generic differences. First, the elimination of the auxiliary fields leads to corrections to the above formula. When the higher-derivative terms are important, these corrections can be significant, drastically modifying the dynamics. The second property is that the equation of motion for the auxiliary field $F^{i}$ of a chiral multiplet is now a cubic equation-- whereas previously it was linear. Thus, in general it admits three distinct solutions, which, after substituting back into the Lagrangian, lead to three inequivalent theories. In this paper, we present the basic properties of each of these three branches.

The bulk of the paper presents our general formalism. It is useful, therefore, to give an explicit example-- which we do by constructing the supergravity extension of a particular DBI action. This allows us to display the specific corrections to both the kinetic and potential terms induced by the elimination of the auxiliary fields when higher-derivative terms are present. We also analyze one of the new branches of the supergravity DBI theory, commenting on the implications of our results for models of DBI inflation. In particular, we find that in the relativistic regime of the DBI theory, the potential automatically becomes negative-- rendering inflation impossible. These findings illustrate the significance that the auxiliary fields can have on the dynamics of a given model.

There are many potential applications of our results, particularly in early universe cosmology. For example, cosmological models that are constructed in-- or inspired by --string theory should admit an effective ${\cal{N}}=1$ supergravity description in four-dimensions. These theories typically have scalar fields arising as the moduli associated with branes \cite{Donagi:1999jp}, flux \cite{Buchbinder:2002ji,Buchbinder:2002pr} or the compactification manifold. For most-- if not all --of these models, whether they are of DBI inflation \cite{Silverstein:2003hf}, $k$-inflation \cite{ArmendarizPicon:1999rj}, $k$-essence \cite{ArmendarizPicon:1999rj}, ekpyrotic/cyclic cosmology \cite{Khoury:2001wf,Steinhardt:2001st,Buchbinder:2007ad,Lehners:2008vx}, effective theories of Galileons \cite{Trodden:2011xh} or higher-derivative induced cosmic bounces \cite{ArkaniHamed:2003uy,Creminelli:2006xe,Lehners:2011kr}, the proper setting is 
supergravity-- and all contain phases where the dynamic description includes scalar higher-derivative terms. We hope to apply our formalism to these models in the future. 

The plan of the paper is the following. We begin in Section  \ref{sectionFlat} by reviewing the construction of higher-derivative kinetic terms for chiral multiplets in global supersymmetry; that is, when gravity is neglected. Then, in Section \ref{sectionX2}, it is shown how this construction can be generalized to supergravity. We proceed by eliminating the auxiliary fields one by one, beginning with $b_{m}$ and $M$ of pure supergravity. . The auxiliary fields $F^i$ of the chiral multiplets require special attention, and Section \ref{sectionauxiliary} is devoted to them. In Section \ref{sectionDBI} we apply our formalism to an example of the DBI action. For the benefit of the reader, we include short summaries of our results at the end of each subsection in \ref{sectionauxiliary} and \ref{sectionDBI} . After concluding in Section \ref{sectionconclusion}, we add Appendices describing the difference of our formalism with the framework of Baumann and Green \cite{Baumann:2011nk,Baumann:2011nm}, as well as comments on K\"{a}hler invariance in the present context. The notation and conventions of the book by J. Wess and J. Bagger \cite{Wess:1992cp} are used throughout the paper. 

\section{Higher-Derivative Chiral Superfields in \\ Flat Superspace} \label{sectionFlat}

We begin by considering global $\mac{N}=1$ supersymmetry in flat four-dimensional spacetime. The associated supersymmetry algebra is given by
\be
\{ Q_\a , \bar{Q}_{\dot\a} \} = -2 \sigma^m_{\a \dot\a} P_m, \label{burt3}
\ee
where $Q_\a,\bar{Q}_{\dot\a}$ and $P_m=-\I \pt_m$ generate supersymmetry and translations respectively. Here $\a,\beta,...$ and $\dot{\a},\dot\beta,...$ are the conjugate indices of two-component Weyl spinors and $m,n...$ are spacetime indices. 
To construct supersymmetric Lagrangians in this context, it is useful to work in flat superspace where, in addition to the four ordinary spacetime dimensions (with coordinates $x^m$), one adds four fermionic, Grassmann-valued dimensions (with coordinates $\th_\a,\bar\th_{\dot\a}$). In terms of these coordinates, the supersymmetric generators are represented by the superspace derivatives
\be
D_\a = \frac{\pt}{\pt\th^\a} + \I \sigma_{\a \dot\a}^m \bar\th^{\dot\a} \pt_m, \qquad \bar{D}_{\dot\a} = - \frac{\pt}{\pt\bar\th^{\dot\a}} - \I \th^\a \sigma_{\a\dot\a}^m \pt_m
\ee
which satisfy the algebra
\be 
\{ D_\a, \bar{D}_{\dot\a} \} = -2\I \sigma^m_{\a\dot\a} \pt_m \ .
\label{mtblanc1}
\ee

Any supermultiplet can be obtained as an expansion of a superfield, appropriately constrained, in the anti-commuting coordinates $\th,\bar\th$. The expansion terminates at order $\th\th\bar\th\bar\th$ because of the Grassmann nature of these coordinates.  
 For example, a chiral superfield $\Phi$, defined by the constraint
 \begin{equation}
\bar{D} \Phi =0 \ ,
 \label{burt1}
 \end{equation}
 has the expansion
\begin{eqnarray}
&&\Phi = A(x) + \sqrt{2} \th\chi(x) + \th\th F(x)   \nonumber \\
&&\quad + \I \th\sigma^m \bar\th \pt_m A(x) - \frac{\I}{\sqrt{2}}\th\th\pt_m \chi(x)\sigma^m \bar\th + \frac{1}{4}\th\th\bar\th\bar\th \Box A(x), \label{burt2}
\end{eqnarray}
where $A$ is a complex scalar, $\chi_\a$ is a spin-$\frac12$ fermion and $F$ is a complex auxiliary field-- which, for Lagrangians with canonical kinetic energy, is not a dynamical degree of freedom. 
$(A,\chi,F)$ are the component fields of the chiral supermultiplet. The component expansion (\ref{burt2}) can be simplified by using the coordinates $y^m = x^m+ \I \th \sigma^m \bar\th,$ in terms of which 
\bea
\Phi = A(y) + \sqrt{2} \th\chi(y) + \th\th F(y).
\eea
This form of the component expansion has a straightforward generalization to curved superspace, as we will see shortly. It also suggests an alternative way of defining component fields, which turns out to be more useful in supergravity. Consider, for example, the chiral supermultiplet $\Phi$. We note that one can also define the components of $\Phi$ as
\bea
A &\equiv& \Phi\mid \\ \chi_\a &\equiv& \frac{1}{\sqrt{2}} D_\a \Phi\mid \\ F &\equiv& -\frac{1}{4} D^2 \Phi\mid 
\eea
where $\mid$ denotes taking the lowest component. It is straightforward to check that these fields are identical to those in the $\th,\bar\th$ expansion \eqref{burt2}.

A general feature of superspace is that the highest component (that is, the $\th\th\bar\th\bar\th$ component) transforms under supersymmetry 
into a total spacetime derivative. Thus, the highest component of a superfield can be used to construct a supersymmetric Lagrangian. Because of the Grassmann nature of the fermionic coordinates, one can isolate the top component by integrating over superspace with $\d^2\th \d^2\bar\th.$ Moreover, one can replace the $\d^2\th \d^2\bar\th$ integral over all superspace by a chiral integral $-\frac14 \d^2 \th \bar{D}^2$ using the chiral projector $\bar{D}^2$. This follows from the flat superspace relation $\bar{D}^3=0.$

In a previous paper \cite{Khoury:2010gb}, it was shown how to construct supersymmetric actions involving higher-derivatives of chiral superfields. The construction is based on a particular supersymmetric extension of the scalar field Lagrangian $(\pt\phi)^4$ given by $D^\a \Phi D_\a \Phi \bar{D}_{\dot\a} \Phid \bar{D}^{\dot\a} \Phid$. Ignoring the fermion $\chi$, this superfield contains only the $\th\th\bar\th\bar\th$ component 
\begin{eqnarray}
&& D^\a \Phi D_\a \Phi \bar{D}_{\dot\a} \Phid \bar{D}^{\dot\a} \Phid  \label{X2}= \th\th\bar\th \bar\th \Big(16 (\pt A)^2 (\pt A^*)^2 \label{X2} \\
&& \qquad \qquad\qquad\qquad\qquad   - 32 \, |\pt A|^2  |F|^2  + 16 |F|^4 \Big) \ , \nonumber
\end{eqnarray} 
where the complex scalar $A$ is composed of two real scalars $\phi,\xi$ as
\be
A = \frac{1}{\sqrt{2}} (\phi + \I \xi)
\ee
and $|\pt A|^2 \equiv \pt A \cdot \pt A^*.$ Thus, the superspace integral of the superfield (\ref{X2}) yields the term 
\be
16 (\pt A)^2 (\pt A^*)^2 = 4 (\pt\phi)^4 + 4 (\pt\xi)^4 - 8 (\pt\phi)^2 (\pt\xi)^2 + 16 (\pt\phi\cdot \pt\xi)^2
\ee
plus terms involving the auxiliary field $F$. Hence, \eqref{X2} constitutes a possible supersymmetric extension of $(\pt\phi)^4.$ The relationship to a different supersymmetric extension of $(\pt\phi)^4$ is discussed in Appendix A. In this paper, we concentrate on (\ref{X2}) since this superfield possesses  several particularly useful properties:
\begin{itemize}
\item It constitutes a supersymmetric extension of the precise expression $(\pt\phi)^4,$ and does not contain other terms involving $\phi$ alone.
\item Despite the higher-derivative nature of the superfield, the auxiliary field $F$ does not obtain a kinetic energy. This is non-trivial, as on dimensional grounds a term such as $|A|^{2} |\pt F|^2$ could have arisen, and implies that $F$ remains truly auxiliary.
\item As pointed out in \cite{Khoury:2010gb}, the auxiliary field now appears at quartic order in the action and, thus, its equation of motion is cubic. Hence, in contrast to the usual two-derivative supersymmetric theories, there exist now up to three different solutions for $F$. We will explore this issue much further in Section \ref{sectionauxiliary}.
\item Finally, the most crucial property for our present purposes is the fact that the bosonic part of $D^\a \Phi D_\a \Phi \bar{D}_{\dot\a} \Phid \bar{D}^{\dot\a} \Phid$, given in (\ref{X2}), only contains a non-zero top $\th\th\bar\th\bar\th$ component-- all lower components vanish. It follows that if one multiplies this superfield with any function $T$ of $\Phi,$ $\Phid$ and (an arbitrary number of) their spacetime derivatives, then the component expansion will be given by (\ref{X2}) times $T|,$ where inside $T|$ the chiral superfield $\Phi$ is simply replaced by its lowest component $A.$ This allows one to easily construct a supersymmetric extension of {\it any} higher-derivative scalar Lagrangian containing $(\pt\phi)^4$ as a factor, simply by performing the replacement $\phi \rightarrow \sqrt{2} A \rightarrow \sqrt{2} \Phi$ in the co-factor.
\end{itemize}
This last property was used in \cite{Khoury:2010gb} to construct a supersymmetric extension of theories with Lagrangian $P(X,\phi),$ where $X \equiv -\frac12 (\pt\phi)^2.$ Specifically, for
\be
P(X,\phi) =  \sum_{n \geq 1} a_n(\phi)X^n
\ee
it was shown that the higher-derivative terms in the supersymmetric generalization are the ${\rm d}^2 \theta{\rm d}^2 {\bar{\theta}}$ integral of
\bea
\frac{1}{16}D \Phi D \Phi \bar{D} \Phi^{\dagger} \bar{D} \Phi^{\dagger} ~T(\Phi,\Phi^{\dagger},\partial_{m}\Phi,\partial_{n}\Phi^{\dagger}),
\label{SusyPofX}
\eea
where 
\begin{eqnarray}
T(\Phi,\Phi^{\dagger},\partial_{m}\Phi,\partial_{n}\Phi^{\dagger})&=& \sum_{n \geq 2} a_n  \left(\frac{1}{32}\{D,\bar{D}\}(\Phi + \Phi^{\dagger})\{D,\bar{D}\}(\Phi + \Phi^{\dagger})\right)^{n-2} \nonumber \\ &=& \sum_{n \geq 2} a_n  \left(\frac{1}{4}\pt^m(\Phi + \Phi^{\dagger})\pt_m(\Phi + \Phi^{\dagger})\right)^{n-2} \ ,
\label{SusyPofX1}
\end{eqnarray}
$a_n =a_n\left(\frac{\Phi + \Phi^{\dagger}}{\sqrt{2}}\right)$ and we have made use of \eqref{mtblanc1} to write $\{D,\bar{D}\} \propto \partial_{m}$.

Particular applications were a supersymmetric form of the DBI action, as well as a supersymmetric ghost condensate theory-- both in flat spacetime. However, the most interesting phenomenological consequences occur when these models are coupled to gravity-- for example, inflation driven by the DBI part of the action or cosmic bounces induced by a ghost condensate.  It is, therefore, of interest to include gravity in the analysis. In a supersymmetric context, this means extending the above construction to {\it curved} superspace. This will be the topic of the next section.

\section{Higher-Derivative Chiral Superfields in \\ Curved Superspace} \label{sectionX2}

We now want to extend the above results to ${\mac{N}}=1$ supergravity, obtained by ``gauging'' the supersymmetry algebra \eqref{burt3}. Loosely speaking, the gauge field associated with the translation generators $P_m$ is the vierbein ${e_m}^a$ (where $a,b,...$ denote tangent space indices), while the gauge field associated with $Q_\a$ is the gravitino $\psi_{m\a}$. As with global supersymmetry, supergravity is most easily expressed in superspace--now, however, with non-vanishing curvature. In this case, one can introduce new fermionic coordinates $\Th$ which are defined precisely so that the $(A, \chi, F)$ components of a chiral superfield $\Phi$ arise as the coefficients of the expansion 
\be
\Phi = A + \sqrt{2} \Th^\a \chi_\a + \Th^\a \Th_\a F.
\ee 
In curved superspace, supersymmetric Lagrangians can be constructed from the chiral integrals
\be
\int \d^2 \Th (\bcD^2 - 8 R) L,
\ee
where $L$ is a scalar, hermitean function. Note that the chiral projector in curved superspace is $\bcD^2 - 8R,$ where $\bcD_{\dot\a}$ is a spinorial component of the curved superspace covariant derivative $\cD_A=(\cD_a,\cD_\a,\bcD_{\dot\a})$ and $R$ is the curvature superfield. In its component expansion, $R$ contains the Ricci scalar $\cR$ and the gravitino $\psi_m,$ as well as the auxiliary fields of 
supergravity-- namely a complex scalar $M$ and a real vector $b_m.$ The purely bosonic components in the $\Th$ expansion of $R$ are
\be
R = -\frac16 M + \Th^2 \big( \frac{1}{12}\cR -\frac19 MM^* - \frac{1}{18} b_m b^m + \frac16 \I {e_a}^m \cD_m b^a\big) \ .
\ee
A second superfield that we will need is the chiral density $\mac{E}$. This contains the determinant of the vierbein $e,$ as well as $M$ and $\psi_m.$ Its bosonic expansion is  
\be
2 \mac{E} = e (1 - \Th^2 M^*) \ .
\ee
For a complete discussion of curved superspace we refer the reader to \cite{Wess:1992cp}, whose notation and formalism we use.
 
In this paper, we will construct a supergravitational extension of a generic higher-derivative scalar field Lagrangian with, however, {\it all fermions set to zero}. We ignore the fermions for two reasons; first, to reduce the complexity of the discussion and, second, so as to emphasize the important physics occurring in the bosonic sector of this theory. The more complete Lagrangian, with all fermions turned on, will be discussed in follow-up papers, where the physics associated with them will be elucidated. As a warm-up, we construct the theory of chiral superfields without higher-derivatives coupled to supergravity -- again with all fermions set to zero. We start by introducing an hermitean  {\it K\"{a}hler potential} $K(\Phi^i,\Phi^{\dagger k*})$ of the chiral 
superfields $\Phi^i$ (where $i=1,2,\dots$ enumerates the fields), along with a holomorphic {\it superpotential} $W(\Phi^i)$. The associated Lagrangian is given by
\bea
\mac{L} &= &\int \d^2\Theta 2\mac{E}\Big[ \frac{3}{8}(\bcD^2-8R) e^{-K(\Phi^i,\Phi^{\dagger k*})/3}+W(\Phi^i)\Big]+h.c. \\
&=& -\frac{3}{32}e\cD^2\bcD^2 e^{-K/3}\mid - \frac{3}{8}eM^*\bcD^2 e^{-K/3}\mid - \frac{1}{8}eM\cD^2 e^{-K/3}\mid \nn \\ 
&& +e\big(- \frac14 \cR -\frac16MM^*  +\frac16 b^a b_a -\frac{\I}{2}{e_a}^m \cD_m b^a\big)e^{-K(A,A^*)/3} \nn \\ && -eW(A)M^* + e \pt W_i F^i + h.c.,
\label{burt4}
\eea
where $\pt W_i = \frac{\pt W}{\pt A^i}$. This Lagrangian is meant to be integrated over spacetime to yield an action. With this in mind, we integrate by parts \footnote{We only use integration by parts on this part of the action, as we will not multiply this with any field dependent factor later in our analysis.} to obtain
\bea
\frac1e \mac{L} &=& e^{-K/3} \big(-\frac12 \cR -\frac13 MM^* +\frac13 b^a b_a\big) \nn \\ && + 3 \Big(\frac{\pt^2 e^{-K/3}}{\pt A^i \pt A^{k*}}\Big) (\pt A^i \cdot \pt A^{k*} - F^i F^{k*}) \nn \\ &&  + \I b^m (\pt_m A^i \frac{\pt e^{-K/3}}{\pt A^i} - \pt_m A^{k*} \frac{\pt e^{-K/3}}{\pt A^{k*}}) + MF^i \frac{\pt e^{-K/3}}{\pt A^i}  \\ && + M^* F^{k*} \frac{\pt e^{-K/3}}{\pt A^{k*}}-WM^* - W^* M + \pt W_i F^i + \pt W^*_{k*} F^{k*} \nn.
\eea

We now add the higher-derivative kinetic terms for the chiral superfields, following the results derived previously in flat superspace \cite{Khoury:2010gb}. As reviewed above in Section \ref{sectionFlat}, the superspace integral of $\cD \Phi \cD \Phi \bcD \Phid \bcD \Phid$ contains the term $16 (\pt A)^2 (\pt A^*)^2$ in its component expansion. Hence, we add such a term to the Lagrangian, now, however, in a  manifestly diffeomorphism invariant manner.\footnote{We thank Ilarion Melnikov for stressing the issue of target space diffeomeorphism invariance to us.} Specifically, we introduce
\bea
\mac{L}_{\rm{h-d}} &=& -\frac18 \int \d^2\Theta 2\mac{E} (\bcD^2-8R) \cD \Phi^i \cD \Phi^j \bcD \Phi^{\dagger k*} \bcD \Phi^{\dagger l*} \, T_{ijk*l*}+h.c.\nn \\ &=& 16 \frac{}{}e(\pt A^i \cdot \pt A^j) (\pt A^{k*}\cdot \pt A^{l*}) \, T_{ijk*l*}| \nn \\ && -32\frac{}{}e F^i F^{k*} (\pt A^j \cdot \pt A^{l*}) \, T_{ijk*l*}| \nn \\ && + 16 eF^i F^j F^{k*} F^{l*} \, T_{ijk*l*}|,
\label{burt5}
\eea
where $T_{ijk*l*}|$ is the lowest component of the tensor superfield $T_{ijk*l*}.$ Let us clarify the meaning of $T_{ijk*l*}$. First, this superfield transforms as a four-index tensor on the K\"{a}hler manifold in which the scalar fields take their values (we know that the target space is a K\"{a}hler manifold from the two-derivative part of the action-- see Appendix B for more details on this point) and, thus, ensures target space diffeomorphism invariance. Second, $T_{ijk*l*}$ is required to be hermitian and symmetric in the pair of indices $i,j$ as well as in $k^*,l^*$. Third, any tensor satisfying these constraints can be multiplied by an arbitrary real function of the chiral superfields and an unlimited number of their ${\cal{D}}_{m}$ covariant derivatives, as long as all indices stemming from the covariant derivatives are contracted. Examples of $T_{ijk*l*}|$ include $\frac12(g_{ik*}g_{jl*}+g_{il*}g_{jk*}),$ where $g_{ij^{*}}$ is the K\"ahler metric, and the Riemann tensor $R_{ik*jl*}.$  However, more general-- non-geometric --choices respecting the required symmetries are equally possible\footnote{In all examples in this paper,  we will, for specificity, choose $T_{ijk*l*}|$ to be proportional to $\frac12(g_{ik*}g_{jl*}+g_{il*}g_{jk*})$.}. The fact that one can multiply this tensor with an arbitrary function of the chiral superfields and their spacetime derivatives means that we can obtain a supergravity extension of any term that involves $(\pt\phi)^4$ as a factor and, thus, by dividing out by $(\pt\phi)^4$ if necessary, of {\it any} higher-derivative scalar Lagrangian. An illustrative example of the usefulness of this property is provided by the DBI action presented in Section \ref{sectionDBI}. 

The new higher-derivative terms necessarily enter with at least one new mass scale, which renders the action dimensionless. In fact, since the $T$ tensor can be composed of many terms, it can contain a number of such masses. In a given application, these mass scales will, of course, be important in determining the significance of the various terms. However, in the present paper, we have set these mass scales to unity-- so as to simplify our formulae and because they are easy to reintroduce.  

The sum of the two actions \eqref{burt4}+\eqref{burt5} does not lead to ordinary Einstein frame gravity but, rather, to a scalar-gravity theory of the form $e^{-K/3}\cR.$ One can transform the action into Einstein frame by performing the Weyl rescaling
\be
{e_n}^a \rightarrow {e_n}^a e^{K/6}.
\ee Note that the higher-derivative term does not contribute to the gravity-scalar coupling and, hence, we can perform the same Weyl rescaling as in ordinary chiral supergravity without higher-derivatives. This is a non-trivial feature of our framework, which greatly facilitates subsequent calculations. Adding the two actions above, and performing the Weyl rescaling, gives
\bea
\frac1e \mac{L}_{\rm{Weyl}} &=& -\frac12 \cR -\frac34 \frac{\pt^m(e^{-K/3})\pt_m(e^{-K/3})}{e^{-2K/3}} + \text{total derivative} \nn \\ &&
+ 3 e^{K/3} \Big(\frac{\pt^2 e^{-K/3}}{\pt A^i \pt A^{k*}}\Big) \pt A^i \cdot \pt A^{k*} \nn \\ && +\frac13 b^a b_a + \I e^{K/3} b^m (\pt_m A^i \frac{\pt e^{-K/3}}{\pt A^i} - \pt_m A^{k*} \frac{\pt e^{-K/3}}{\pt A^{k*}}) \nn \\ &&
- 3 e^{2K/3}\Big( \frac{\pt^2 e^{-K/3}}{\pt A^i \pt A^{k*}}\Big) F^i F^{k*}  \nn \\ && + e^{2K/3} MF^i\Big( \frac{\pt e^{-K/3}}{\pt A^i}\Big) + e^{2K/3} M^* F^{k*}\Big( \frac{\pt e^{-K/3}}{\pt A^{k*}}\Big) \nn \\ && -\frac13 e^{K/3} MM^* -e^{2K/3} WM^* - e^{2K/3}W^* M \nn \\ && + \frac{}{}e^{2K/3}\pt W_i F^i + e^{2K/3}\pt W^*_{k*} F^{k*} \nn \\ &&
+ 16 \frac{}{}(\pt A^i \cdot \pt A^j) (\pt A^{k*}\cdot \pt A^{l*}) \, T_{ijk*l*\rm{Weyl}}| \nn \\ && -32 \frac{}{}e^{K/3} F^i F^{k*} (\pt A^j \cdot \pt A^{l*}) \, T_{ijk*l*\rm{Weyl}}| \nn \\ && + 16 e^{2K/3} F^i F^j F^{k*} F^{l*} \, T_{ijk*l*\rm{Weyl}}|.
\eea
The subscript ``Weyl'' on $T_{ijk*l*\rm{Weyl}}|$ indicates that if this expression involves the spacetime metric, then it must be rescaled as $g_{mn} \rightarrow e^{K/3} g_{mn}$. Henceforth, we drop the total derivative term. To proceed, we want to eliminate the auxiliary fields. We begin with $b_m$, whose equation of motion does not involve the higher-derivative terms and is given by
\be
b_m = \frac{\I}{2} (\pt_m A^i K_{,A^i} - \pt_m A^{k*} K_{,A^{k*}}) \ . 
\ee 
Substituting this back into the action, while also defining 
\begin{equation}
N = M + K_{,A^{k*}} F^{k*} \ ,
\label{burt6}
\end{equation}
yields
\bea
\frac1e \mac{L}_{\rm{Weyl}} &=& -\frac12 \cR - g_{ik*} \pt A^i \cdot \pt A^{k*} + g_{ik*} e^{K/3} F^i F^{k*} -\frac13 e^{K/3} NN^* \nn \\ && + \frac{}{}e^{2K/3}\big(-WN^* - W^* N + F^i (D_A W)_i + F^{k*} (D_A W)^*_{k*}\big) \nn \\ && 
+ \frac{}{}16 (\pt A^i \cdot \pt A^j) (\pt A^{k*}\cdot \pt A^{l*}) \, T_{ijk*l*\rm{Weyl}}| \nn \\ && -32 \frac{}{}e^{K/3} F^i F^{k*} (\pt A^j \cdot \pt A^{l*}) \, T_{ijk*l*\rm{Weyl}}| \nn \\ && + 16 e^{2K/3} F^i F^j F^{k*} F^{l*} \, T_{ijk*l*\rm{Weyl}}|, \label{LagrangianWithF}
\eea
where the K\"{a}hler metric is $g_{ik*} = \frac{\pt^2 K}{\pt A^i \pt A^{k*}}$ and $D_A W_i = \pt W_i + K_{,A^i} W$ is the K\"{a}hler covariant derivative.  The equation of motion for $N$ is again independent of the higher-derivative terms, and is simply  
\be
N = -3 e^{K/3} W \ .
\ee
Plugging this back into the action gives
\bea
\frac1e \mac{L}_{\rm{Weyl}} &=& -\frac12 \cR - g_{ik*} \pt A^i \cdot \pt A^{k*} + g_{ik*} e^{K/3} F^i F^{k*}  \nn \\ && +\frac{}{} e^{2K/3}[F^i (D_A W)_i + F^{k*} (D_A W)^*_{k*}]  + 3 e^K WW^* \nn \\ && 
+ \frac{}{}16 (\pt A^i \cdot \pt A^j) (\pt A^{k*}\cdot \pt A^{l*}) \, T_{ijk*l*\rm{Weyl}}| \nn \\ && -32 \frac{}{}e^{K/3} F^i F^{k*} (\pt A^j \cdot \pt A^{l*}) \, T_{ijk*l*\rm{Weyl}}| \nn \\ && + 16 e^{2K/3} F^i F^j F^{k*} F^{l*} \, T_{ijk*l*\rm{Weyl}}|. \label{EqActionWithF}
\eea

In the next section, we will discuss the remaining auxiliary field, namely $F$, in great detail. Before doing so, however, let us write out-- for completeness --the supersymmetry transformations of the above theory. As everywhere in this paper, we only consider the bosonic contributions and, hence, the fermionic variations only. The original transformations are given by
\bea
\delta_\epsilon \chi^i &=& \I \sqrt{2} \sigma^m \bar{\epsilon} \pt_m A^i + \sqrt{2} \epsilon F^i, \\
\delta_\epsilon \psi_m &=& -2\cD_m \epsilon + \I {e_m}^a \left(\frac13 M \sigma_a\bar{\epsilon} + b_a \epsilon + \frac13 b^c \epsilon \sigma_c\bar\sigma_a \right),
\eea
where the supersymmetry parameter is denoted by $\epsilon.$ Weyl rescaling is performed via
\bea
\chi &\rightarrow& e^{-K/12} \chi \ , \\ \psi_m &\rightarrow& e^{K/12} \psi_m \ , \\ \epsilon &\rightarrow& e^{K/12} \epsilon \ .
\eea
As discussed in \cite{Wess:1992cp}, the gravitino must also be shifted as
\bea 
\psi_m \rightarrow \psi_m + \I \frac{\sqrt{2}}{6}K_{,A^{k*}}\bar{\chi}^{k*}
\eea
in order for the fermionic kinetic terms to be in canonical form. Plugging in the solutions for $M$ and $b_m$, we obtain
\bea
\delta_\epsilon \chi^i &=& \I \sqrt{2} \sigma^m \bar{\epsilon} \pt_m A^i + \sqrt{2} e^{K/6} \epsilon F^i, \label{EqVarChi} \\ 
\delta_\epsilon \psi_m &=& 2 \big(\cD_m + \frac14 (K_{,A^i}\pt_m A^i-K_{,A^{k*}}\pt_m A^{k*})\big)  \epsilon+ \I e^{K/2}W\sigma_m \bar{\epsilon}.
\eea
Note that, although $M$ depends on $F$ via its definition in terms of $N,$ the shift of the gravitino subsequently removes the $F$ dependence from the gravitino variation. In (\ref{EqVarChi}), however, $F$  will have to be replaced by the particular solution for $F$ under consideration. It is to these solutions that we now turn our attention.

\section{The Auxiliary Field $F$} \label{sectionauxiliary}

We now consider the most interesting of the auxiliary fields, namely $F.$ Three remarks are in order. First, despite the fact that we have added higher-derivative terms, $F$ does not obtain a kinetic term in our formalism. This is non-trivial in this context, and implies that $F$ remains a truly auxiliary field. Second, there is some subtlety regarding the quantum theory associated with this action. For standard two-derivative actions, where $F$ only appears at quadratic order, we can do one of two equivalent things: either eliminate $F$ using its algebraic equation of motion, or, in the path integral formalism, simply integrate over $F.$ This second approach leads to a Gaussian integral, and the end result is the same as eliminating $F$ via its equation of motion. In the higher-derivative formalism presented in this paper, since $F$ now appears at fourth order, this equivalence is no longer preserved. Thus, there is some ambiguity as to what the correct quantum theory should be. Since, in this paper, we are only studying the theory at the classical level, we will proceed by eliminating $F$ via its equation of motion. 
This brings us to our third remark. The equation of motion for $F$ is easily derived from the action (\ref{LagrangianWithF}) and reads
\be
g_{ik*} F^i+ e^{K/3} (D_A W)^*_{k*} + 32 F^i (e^{K/3} F^j F^{l*} - \pt A^j \cdot \pt A^{l*}) T_{ijk*l*\rm{Weyl}}| = 0. 
\ee
This equation is now cubic in $F$ and, thus, it can have up to three inequivalent solutions. As we will see, these different solutions lead to {\it different} theories! 
From now on, we will restrict our analysis to a single chiral superfield $\Phi^1 = \Phi$, the extension to multiple superfields being straightforward to implement. In this case, the equation of motion for $F$ becomes
\be
K_{,AA^*} F+ e^{K/3} (D_A W)^* + 32 F (e^{K/3} |F|^2 - |\pt A|^2) \mac{T} = 0, \label{EqofMotionF}
\ee
where
\be
|\pt A|^2 = \pt A \cdot \pt A^* = g^{mn}\pt_m A \pt_n A^*
\ee
and where we use the simplified notation
\be
\mac{T} \equiv T_{111*1*\rm{Weyl}}|. \label{ScriptT}
\ee
Note that $\mac{T}$ is effectively an arbitrary real scalar function of $A,A^*$ and their spacetime covariant derivatives $\cD_m \dots \pt_n A,$ $\cD_m \dots \pt_n A^*.$

To proceed, let us first consider the case where the superpotential is absent. The effect of turning on a superpotential will be discussed thereafter.

\subsection{Without A Superpotential} \label{sectionnoW}

We first analyze the case with vanishing  superpotential, $W=0.$ The equation for $F$ then becomes
\be
F \big(K_{,AA^*} + 32 \mac{T}(e^{K/3} |F|^2 - |\pt A|^2)\big) = 0. \label{FFFstarEquation}
\ee
This has {\it two} solutions, which we denote by $F_0$ and $F_{\rm{new}}$ respectively. The first solution is the trivial one, where $F_0 = 0$.  In this case, the Lagrangian becomes purely kinetic, as expected, and is given by
\bea
\frac1e \mac{L}_{W=0,F_{0}=0} &=& -\frac12 \cR - K_{,AA^*} |\pt A|^2  
+ 16 (\pt A)^2 (\pt A^{*})^2 \, \mac{T} .
\eea
However, there is a second-- non-trivial --solution corresponding to the large bracket in (\ref{FFFstarEquation}) vanishing; that is,
\be
|F_{\rm{new}}|^2 = -\frac{1}{32\, \mac{T}}e^{-K/3} K_{,AA*} + e^{-K/3} \, |\pt A|^2. \label{Fnew}
\ee
Putting this equation into \eqref{EqActionWithF}, the Lagrangian becomes
\bea
\frac1e \mac{L}_{W=0,F_{\rm{new}}} &=& - \frac12 \cR + 16 \mac{T} \big((\pt A)^2(\pt A^*)^2 - (\pt A \cdot \pt A^*)^2\big) \nn \\ && - \frac{1}{64 \mac{T}}(K_{,AA^*})^2. \label{WzeroNewBranch1}
\eea
Note that this theory is not continuously connected to the ordinary two-derivative supergravity since in the limit $\mac{T} \rightarrow 0$ the term proportional to $1/{\cal{T}}$ blows up. 
Remarkably, the ordinary kinetic term has vanished-- being replaced by purely higher-derivative terms! In making this statement, we have discarded one special case: since $\mac{T}$ is arbitrary in our formalism, there is the possibility that an ordinary kinetic term could arise from a particular form of $\mac{T},$ such as $\mac{T} \supset -|\pt A|^2/\big((\pt A)^2(\pt A^*)^2 - (\pt A \cdot \pt A^*)^2\big).$ We will, in fact, examine such a situation in Section \ref{sectionDBI}. However, for now, let us proceed with the case where $\mac{T}$ is a function of the fields $A,A^*$ only, without derivatives. 

Then, something interesting occurs. Although we have set the superpotential to zero in the present section, the elimination of $F$ in this new branch leads to a non-vanishing potential energy given by
\be
V_{\rm{new}}=\frac{1}{64 \mac{T}}(K_{,AA^*})^2.
\label{burt7}
\ee
This can be positive or negative, depending on the sign of the tensor $\mac{T}.$ The form of the potential depends on which K\"{a}hler potential and which $\mac{T}$ tensor one considers. This choice is largely unrestricted, but there is one consistency condition that must be satisfied; that is,  the right-hand side of \eqref{Fnew} must be positive.  This can be achieved in one of two ways, which we examine in turn-- 1) either $K_{,AA^*} \mac{T}<0$ and $\langle \pt A \rangle$ is small, or 2) at least one of the two real scalars that make up $A$ must have large spatial gradients. 

In the first case, where the scalars do {\it not} have large spatial gradients, it is clear that one must take $\mac{T}$ negative when the K\"{a}hler metric has the usual positive sign. It follows that the potential \eqref{burt7} is negative.
The second case corresponds to the situation where some spatial gradients are {\it large}. To explore this, write the complex scalar $A$ in terms of two real scalars $\phi,\xi$ as
\be
A = \frac{1}{\sqrt{2}} (\phi + \I \xi).
\ee
We will choose the $\mac{T}$ tensor to be of the canonical form $(K_{,AA^*})^2 $, but allow for an additional real multiplicative factor $v(\phi,\xi)$. That is, take
\be
\mac{T}=(K_{,AA^*})^2 v(\phi,\xi).
\ee 
Then, in a flat Robertson-Walker background with metric $\d s^2 = - \d t^2 + a(t)^2 \d \bf{x}^2,$ the action becomes
\bea
\int \d^4x \mac{L}_{W=0,F_{\rm{new}}} &=& \int \d^4 x a^3 \Big(- 3 \frac{\dot{a}^2}{a^2} + \frac{16}{a^2} v(\phi,\xi) (\xi_{,i}^2\dot\phi^2 + \phi_{,i}^2\dot\xi^2 -2\phi_{,i}\xi_{,i}\dot\phi \dot\xi)  \nn \\ && \quad + \frac{16}{a^4} v(\phi,\xi)(\phi_{,i}\xi_{,i}\phi_{,j}\xi_{,j}-\phi_{,i}^2 \xi_{,j}^2)  - \frac{1}{64 v(\phi,\xi)}\Big).
\eea
Even though we have a purely higher-derivative theory in (\ref{WzeroNewBranch1}), one can now see that, via their interactions, the scalars can generate ``ordinary'' kinetic terms for each other. Suppose, for example, that $\xi$ develops a non-trivial spatial profile $\xi = \xi(x^i).$\footnote{Here, we simply assume that such solutions exist. Of course, this has to be verified for any given function $v(\phi,\xi).$ In the case where $v$ depends on $\phi$ alone, for example, there exist solutions where $\phi$ is purely time-dependent and $\xi = a_i x^i$ for some constants $a_i.$} Then the theory becomes  
\bea
&&\int \d^4x \mac{L}_{W=0,F_{\rm{new}}} = \int \d^4 x a^3 \big(- 3 \frac{\dot{a}^2}{a^2}+ \frac{16}{a^2} v(\phi,\xi) (\xi_{,i}^2 (\dot\phi^2-\frac{1}{a^2}\phi_{,j}^2) \nn \\ &&\qquad\qquad\qquad \qquad ~~~~~ + \frac{1}{a^2}\phi_{,i}\xi_{,i}\phi_{,j}\xi_{,j}  -V_{\rm{new}}(\phi,\xi)\big) \ ,
\eea
where 
\bea
V_{\rm{new}} = \frac{1}{64 v(\phi,\xi)}. 
\eea
Because of the additional $\phi_{,i}\xi_{,i}\phi_{,j}\xi_{,j}$ term, the dispersion relation for $\phi$ will be slightly unusual in this background, and one may expect that $\phi$ will develop gradient instabilities over sufficiently long timescales. Be that as it may, the theory does have a very interesting feature; namely, if we require that the kinetic term for $\phi$ be ghost-free, then one must impose the condition that $v(\phi,\xi) > 0.$ This then leads to a {\it positive} potential $V_{\rm{new}}$! In other words, having one of the scalars develop large spatial gradients leads to both a two-derivative kinetic term for the other scalar and a positive potential. In a supergravity context, this property is most unusual and deserves further attention.

\vspace{.5cm}
\noindent{\underline{Summary}:} {\it In the absence of a superpotential, there are two types of solutions for the auxiliary field $F.$ The first is the trivial solution $F=0$. Its substitution leads to a purely kinetic Lagrangian including higher-derivative kinetic terms. However, there exist new solutions $F_{\rm{new}}$ as well. These generate a ``potential without a superpotential''. When the scalar fields develop large spatial gradients, this potential can be positive.}

\subsection{With A Superpotential}

Now introduce a non-vanishing superpotential, and consider the solutions for $F$ in its presence. Multiplying (\ref{EqofMotionF}) with $F^{*}$ shows that $(D_A W)^* F^{*}$ must be real. Thus, one can relate $F$ and $F^*$ via
\be 
F^* = \frac{D_A W}{(D_A W)^*} F
\ee
as long as $D_A W \neq 0,$ which we now assume. One can use this relation to obtain a cubic equation for $F$ alone. This is given by
\be
K_{,AA^*}F + e^{K/3} (D_A W)^* + 32 (e^{K/3}\frac{D_A W}{(D_A W)^*} F^3 - |\pt A|^2 F) \mac{T} =0. \label{FFFEquation}
\ee
In general, this equation admits {\it three} distinct solutions-- which we denote by 
$F_1,F_2,F_3$ --leading to three different theories. One can find these solutions using Cardano's formula. Define
\bea
p &=& e^{-K/3} \frac{(D_A W)^*}{D_A W} \left( \frac{K_{,AA^*}}{32 \mac{T}} -|\pt A|^2 \right),  \\
q &=& \frac{1}{32 \mac{T}} \frac{(D_A W)^{*2}}{D_A W},  \\
D &=& \left(\frac{q}{2}\right)^2 + \left(\frac{p}{3}\right)^3  \\ &=& \frac{1}{(64 \mac{T})^2} \frac{(D_A W)^{*4}}{(D_A W)^2} + \frac{1}{27e^{K}} \frac{(D_A W)^{*3}}{(D_A W)^3} \left( \frac{K_{,AA^*}}{32 \mac{T}} -|\pt A|^2 \right)^3 \ . \nn
\eea
Then the solutions are given by
\be
F_{k+1} = \omega^k F_+ + \omega^{-k} F_- \ ,
\ee
where $k=0,1,2$, $\omega = e^{2\pi\I/3}= -\frac{1}{2} +\I \frac{\sqrt{3}}{2}$ is a cube root of unity and
\be
F_+ = (-\frac{q}{2} + D^{1/2})^{1/3}, \qquad F_- = (-\frac{q}{2} - D^{1/2})^{1/3}.
\ee
The three solutions can also be written as
\bea
F_1 &=& F_+ + F_-, \\
F_2 &=& -\frac{1}{2}(F_+ + F_-) + \I \frac{\sqrt{3}}{2}(F_+ - F_-) ,\\
F_3 &=& -\frac{1}{2}(F_+ + F_-) - \I \frac{\sqrt{3}}{2}(F_+ - F_-) .
\eea
Substituting these back into the action generates three different branches of the theory. We call the theory that results from substituting $F_1$ the {\it ordinary branch}, and the ones associated with $F_2$ and $F_3$ the {\it new branches}, for reasons that will become clear. In general, the solutions presented above are rather complicated. However, to get some insight one can analyze them in different simplifying limits.

\vspace{.5cm}
\noindent{\underline{Summary}:} {\it When a superpotential is present, the auxiliary field $F$ admits three distinct solutions, which lead to three distinct theories. One of these solutions, which we call the ordinary branch, is related to the usual solution for $F$ that one obtains in two-derivative chiral supergravity, while the other two solutions correspond to new branches of the theory.}

\subsubsection{Small Higher-Derivative Terms}
\label{sectionsmallT}

The higher-derivative terms are all proportional to the $\mac{T}$ tensor. Therefore,  by assuming that $\mac{T}$ contains a factor that can be tuned to be small, one can treat such terms as sub-leading. The $\mac{T}\rightarrow 0$ limit then corresponds to $q \ll p^{3/2}$, and gives rise to the approximate expressions
\be
F_\pm^3 = \pm D^{1/2} - \frac{q}{2} = (\frac{p}{3})^{3/2} \Big(\pm 1 - \frac{q}{2}(\frac{3}{p})^{3/2} \pm \frac{27q^2}{8p^3} + {\cal O}(\frac{q^4}{p^{6}})\Big).
\ee
For the ordinary branch, this implies that 
\be
F_1 = -\frac{q}{p} + \frac{q^3}{p^4}  +{\cal O}(\frac{q^4}{p^{9/2}}),
\ee
or, more explicitly,
\bea
F_1 &=& -K^{,AA^*}e^{K/3}(D_A W)^* \nn \\ && +32 \mac{T} e^{4K/3} (K^{,AA^*})^4 (D_A W)^{*2} D_A W \nn \\ && - 32 \mac{T} e^{K/3} (K^{,AA^*})^2 (D_A W)^*|\pt A|^2 + {\cal{O}}({\cal{T}}^{2}) \ .
\eea
Note that this corresponds to a small correction to the usual solution for the auxiliary field $F$ in the presence of a superpotential. Correspondingly, we obtain small corrections in the Lagrangian by substituting this solution for $F$.  To first order in the higher-derivative terms, the Lagrangian becomes
\bea
&& \frac1e \mac{L}_{\rm{ordinary,\mac{T}\rightarrow 0}} = -\frac12 \cR - K_{,AA^*} |\pt A|^2  -\frac{}{}e^K (K^{,AA^*} |D_A W|^2 - 3 |W|^2) \nn \\ && \qquad\qquad\qquad~-32 \frac{}{}e^{K} K^{,AA^*} |D_A W|^2  K^{,AA^*}|\pt A|^2 \, \mac{T} \nn \\ && \qquad\qquad\qquad~+ 16 \frac{}{}(\pt A)^2 (\pt A^{*})^2 \, \mac{T} \nn \\ && \qquad\qquad\qquad~ + 16 e^{2K} (K^{,AA^*}|D_A W|^2)^2 \, (K^{,AA^*})^2 \mac{T} \ .
\eea
An interesting feature is that both the kinetic terms and the potential get corrected. The potential now becomes
\bea
V &=& \frac{}{}e^K (K^{,AA^*} |D_A W|^2 - 3 |W|^2)  \nn \\ && 
  - 16 (e^{K} K^{,AA^*} |D_A W|^2)^2 \, (K^{,AA^*})^2 \mac{T}_{\rm{no \, der.}}, \label{NewPotential}
\eea
where $\mac{T}_{\rm{no \, der.}}$ stands for the part of $\mac{T}$ that does not contain spacetime derivatives. Note that all the correction terms in the Lagrangian above are invariant under K\"{a}hler transformations. 

As an example, consider the case where $K=\Phi \Phid$, $\mac{T}=\tau (K_{,AA^*})^2$ is of canonical form with $\tau$ a small parameter and $W=\Phi^n$, for some positve integer $n$. Then the potential, to first order in $\tau$, is given by $V = \bar{V} + \delta V$ where
\bea
\bar{V} &=& e^{AA^*}(|A|^{2n+2} + (2n-3) |A|^{2n} + n^2 |A|^{2n-2})
\eea
while
\bea
\delta V &=& -16 \tau e^{2AA^*} |A|^{4n-4}(|A|^2 + n)^4.
\eea
At sufficiently large values of $|A|$, the correction term always becomes larger than the original potential-- indicating that our approximation breaks down. However, for small values of $|A|$ the corrections can be trusted. They are typically small, but in certain cases can lead to novel effects. In particular, consider the case where $n=1$; that is, $W=\Phi.$ Then, near the minimum at $A=0$ the potential can be approximated by
\bea
\bar{V}_{n=1} &\approx& 1 + \frac12 |A|^4 + \cdots \ .
\eea  
Note that the $|A|^2=\phi^2 + \xi^2$ term cancels in the expansion. Therefore,  this potential is very flat near the origin, rising only quartically as $(\phi^2 + \xi^2)^2$. The leading order correction to this potential is given by
\bea
\delta V_{n=1} &\approx& -16 \tau (1 + 6 |A|^2 + 16 |A|^4 + \cdots).
\eea
For $\frac{1}{128}>\tau>0,$ the minimum at $A=0$ becomes a local maximum. The potential is now minimized along a circle defined by $|A|^2=12\tau/(1-128\tau).$ In other words, the potential changes from a slowly rising quartic potential with a minimum at the origin to a ``Mexican hat''.

In the limit where the higher-derivative terms are small, the new branches behave very differently. Using the same approximations as above, the $F_{2,3}$ solutions to the auxiliary field equation of motion can be approximated by
\bea
F_2 &=& \frac{\I}{4\sqrt{2}}e^{-K/6}\left( \frac{(D_A W)^* K_{,AA^*}}{(D_A W) \mac{T}} \right)^{1/2} \nn \\ && + \frac{1}{2}K^{,AA^*}e^{K/3} (D_A W)^*  + {\cal O}(\mac{T}^{1/2}), \\
F_3 &=& -\frac{\I}{4\sqrt{2}}e^{-K/6}\left( \frac{(D_A W)^* K_{,AA^*}}{(D_A W) \mac{T}} \right)^{1/2} \nn \\ && + \frac{1}{2}K^{,AA^*}e^{K/3} (D_A W)^* + {\cal O}(\mac{T}^{1/2}). 
\eea
When substituted into the Lagrangian they give, to sub-leading order in $\mac{T}$,
\bea
\frac1e \mac{L}_{\rm{new,\mac{T}\rightarrow 0}} &=& -\frac12 \cR - 2 K_{,AA^*} |\pt A|^2  \nn \\ &&-e^K \big(-\frac32 K^{,AA^*} |D_A W|^2 - 3 |W|^2\big) \nn \\ && +\frac{3}{64 \mac{T}}(K_{,AA^*})^2 .
\eea
Not only do the ordinary kinetic term and the ordinary part of the potential come out with unusual coefficients, but the last term, which is the dominant term in the $\mac{T} \rightarrow 0$ limit, blows up as the higher-derivative terms are made small. This term, which includes the new contribution to the potential
\be
V_{\rm{new}} = -\frac{3}{64 \mac{T}_{\rm{no \, der.}}}(K_{,AA^*})^2,
\ee
shows explicitly that the new branches are separated from the ordinary supergravity theory by an infinite potential barrier (and implies, incidentally, that the scale of supersymmetry breaking will tend to be large in the new branches). The implication is that these new theories cannot be reached dynamically from the ordinary, two-derivative supergravity in the perturbative regime. In other words, one cannot start gradually turning on the higher-derivative terms and end up in one of the new branches. This leaves open the possibility that these branches might be connected to each other when the higher-derivative terms are large. We will explore this limit next.

\vspace{.5cm}
\noindent{\underline{Summary}:} {\it When our higher-derivative chiral supergravity terms are small, then, in the ordinary branch, they lead to correspondingly small corrections to the two-derivative and potential terms via substitution of the auxiliary field $F.$ The new potential is given by 
\bea
V &=& \frac{}{}e^K (K^{,AA^*} |D_A W|^2 - 3 |W|^2)  \nn \\ && 
  - 16 (e^{K} K^{,AA^*} |D_A W|^2)^2 \, (K^{,AA^*})^2 \mac{T}_{\rm{no \, der.}}, 
\eea
where $\mac{T}_{\rm{no \, der.}}$ stands for the part of $\mac{T}$ that does not contain spacetime derivatives.
In the new branches, even small higher-derivative terms lead to drastic changes in the Lagrangian via substitution of the auxiliary field. In this case, the kinetic and potential terms have unusual coefficients, and a large additional (positive or negative) potential 
\be
V_{\rm{new}} = -\frac{3}{64 \mac{T}_{\rm{no \, der.}}}(K_{,AA^*})^2
\ee
is generated.}

\subsubsection{Large Higher-Derivative Terms} \label{sectionlargeT}

We now consider the opposite limit, where the higher-derivative terms are large compared to the ordinary kinetic terms. We can use the approximate expressions
\bea
F_+ + F_- &=& -\frac{q}{p} + {\cal O}(\frac{q^3}{p^4}), \\
F_+ - F_- &=& 2 (\frac{p}{3})^{1/2} + \frac{\sqrt{3}}{4}\frac{q^2}{p^{5/2}} + {\cal O}(\frac{q^4}{p^{9/2}}).
\eea
In the large $\mac{T}$ limit, the ordinary branch solution then becomes
\be
F_1=0 + e^{K/3} (D_A W)^* \frac{1}{32\mac{T}|\pt A|^2}+{\cal O}(\frac1{\mac{T}^2}).
\ee
Substituting this solution into the Lagrangian, we find to subleading order that
\bea
\frac1e \mac{L}_{\rm{ordinary},\mac{T}\rightarrow \infty} &=& -\frac12 \cR - K_{,AA^*} |\pt A|^2 + 3 e^K |W|^2 \nn \\ && 
+ 16 (\pt A)^2 (\pt A^{*})^2 \, \mac{T} .
\eea
The higher-derivative kinetic term, of course, dominates in this limit. Interestingly, the associated potential given by
\be
V_{\mac{T}\rightarrow \infty}=-3 e^K |W|^2 \label{PotentialLargeT}
\ee
is always {\it negative}. This is because-- in the ordinary branch --the auxiliary field $F$ is essentially irrelevant in the limit of large kinetic terms.

For the new branches, the solutions are slightly more involved. They are given by
\bea
F_2 &=& - e^{-K/6}\left(\frac{D_A W^*}{D_A W} \right)^{1/2}|\pt A| \nn \\ && +\frac{1}{64 \mac{T}}e^{-K/6}K_{,AA^*}\left( \frac{D_A W^*}{D_A W} \right)^{1/2}\frac{1}{|\pt A|} \nn \\ && - \frac{1}{64 \mac{T}}e^{K/3} (D_A W)^* \frac{1}{|\pt A|^2} +{\cal O}(\frac1{\mac{T}^2}) \ ,\\
F_3 &=& e^{-K/6}\left( \frac{D_A W^*}{D_A W}\right)^{1/2} |\pt A| \nn \\ && -\frac{1}{64 \mac{T}}e^{-K/6}K_{,AA^*}\left( \frac{D_A W^*}{D_A W} \right)^{1/2}\frac{1}{|\pt A|} \nn \\ && - \frac{1}{64 \mac{T}}e^{K/3} (D_A W)^* \frac{1}{|\pt A|^2} +{\cal O}(\frac1{\mac{T}^2}) \ .
\eea
For the two new branches, to subleading order, the Lagrangian approaches the same large $\mac{T}$ limit 
\bea
\frac1e \mac{L}_{\rm{new},\mac{T}\rightarrow \infty} &=& -\frac12 \cR - 4K_{,AA^*} |\pt A|^2 + 3 e^K |W|^2 \\ && 
+ 16 [(\pt A)^2 (\pt A^{*})^2 - |\pt A|^4] \, \mac{T} . \nn
\eea
The elimination of the auxiliary fields leads to the presence of additional higher-derivative terms, which are of the same order in derivatives as the original ones considered. Furthermore, the normalization of the ordinary kinetic term is changed, while the potential energy, just as for the ordinary branch, has become equal to
(\ref{PotentialLargeT}) and, thus, is also always negative.

Note that in this large $\mac{T}$ limit, the ordinary and new branches are still different. This leads us to conclude that these branches really correspond to entirely separate, and different, theories. It will be interesting to further explore the physical relevance of the new branches. We leave this topic for future work, and only add one comment. The equation of motion for $F$ (\ref{EqofMotionF}) implies that, for the new branches, one must have $\pt A \cdot \pt A^* > 0$ in the large $\mac{T}$ limit. Then, loosely speaking, the sum of spatial gradients in the scalar fields must be larger than their time gradients. It would be interesting to see how this constraint gets implemented by the dynamics, in a situation where the higher-derivative terms come to dominate progressively.

\vspace{.5cm}
\noindent{\underline{Summary}:} {\it When the higher-derivative terms are large, then in both the ordinary and the new branches the potential is given by 
\be
V_{\mac{T}\rightarrow \infty}=-3 e^K |W|^2
\ee
and, hence, is always negative.
This result is of particular significance for cosmological applications.
Furthermore, in the new branches, additional higher-derivative terms are generated via substitution of the auxiliary field. Both in the limit of small and large higher-derivative terms, the new branches are considerably different than the ordinary branch and represent new theories that are not continuously related to it.}

\section{An Example: DBI in Supergravity} \label{sectionDBI}

One can use our formalism to construct a minimal supergravity version of the Dirac-Born-Infeld (DBI) brane action, whose general form includes the bosonic term
\be
S = - \int \d^d x \sqrt{-g} \frac{1}{f(\phi^k)} \left( \sqrt{\det(g_{mn}+f(\phi^k) \pt_m \phi^i \pt_n \phi^j g_{ij}+ \cF_{mn})} -1 \right).
\ee
Here $\cF_{mn}$ represents the field strengths of p-form fields, which we ignore in the present paper. The $\phi^i$ are real scalar fields specifying the position of the brane in the transverse dimensions. The field space metric $g_{ij}$ as well as the (real and positive) function $f(\phi^i)$ arise from both the higher-dimensional metric and the dilaton. DBI actions are well-motivated from string theory, where they arise as the effective actions of both D- and M5-branes \cite{Leigh:1989jq}. Since these branes are of central importance in string theory, it is of interest to study their realizations in supergravity. Moreover, bosonic DBI actions have been used to construct models of inflation with unusual, but interesting, properties. Specifically, because of their higher-derivative terms, they can lead to inflation on potentials that would otherwise be too steep. Additionally, they have characteristic observational predictions, such as equilateral non-Gaussianity in the spectrum of fluctuations--see \cite{Silverstein:2003hf,Langlois:2008qf}. 

The detailed form of the supergravity DBI action will depend on the context. 
Here, we construct the supergravity version of one specific, but illustrative, example; namely, the DBI action derived by Rocek and Tseytlin as the flat superspace effective action of a $D3$-brane in 6 dimensions \cite{Rocek:1997hi}. This action contains two scalars $\phi,\xi$ describing the position of the brane in the two dimensions transverse to the four-dimensional worldvolume. The Lagrangian is given by
\be
\frac1e {\cal L}_{\rm{brane}} = - \frac{1}{f} \left( \sqrt{{\det}(g_{mn} + f \, \pt_m \phi \pt_n \phi + f\, \pt_m \xi \pt_n \xi)} -1 \right)  ,
\ee
where $f=f(\phi,\xi)$ is a real, positive function. It is useful to combine the two real scalars into a complex scalar $A = \frac{1}{\sqrt{2}}(\phi + \I \xi)$ and to re-write the Lagrangian as 
\bea
\frac1e {\cal L}_{\rm{brane}} &=& - \frac{1}{f(A,A^*)} \left( \sqrt{{\det}(g_{mn} + f(A,A^*)\, \pt_m A \pt_n A^*)} -1 \right) \nn \\ 
&=& -\frac{1}{f} \left( \sqrt{1 + 2 f \, |\pt A|^2 + f^2 \, |\pt A|^4 - f^2 \, (\pt A)^2 (\pt A^*)^2} -1 \right)  \\ 
&=& - |\pt A|^2 + \frac{f\, (\pt A)^2 (\pt A^*)^2}{1 + f\, |\pt A|^2+ \sqrt{(1 + f\, |\pt A|^2)^2 - f^2 \, (\pt A)^2 (\pt A^*)^2}}. \nn
\eea
This action is in a form perfectly suited to our framework. Comparing with Eqs. (\ref{EqActionWithF}) and (\ref{ScriptT}), we can see that one should take $K=\Phi \Phid$ and choose
\bea
16 T_{111*1*\rm{Weyl}}| &\equiv& 16 \mac{T}_{\rm{DBI}}  \\ &=& \frac{f}{1 + f\, |\pt A|^2  + \sqrt{(1 + f\, |\pt A|^2)^2 - f^2 \, (\pt A)^2 (\pt A^*)^2}}. \nn
\eea 
It is then straightforward to write out the curved superspace version of this DBI Lagrangian. It is given by
\bea
{\cal L}_{\rm{DBI}}  &=& \int \d^2\Theta 2\mac{E}\Big[ \frac{3}{8}(\bcD^2-8R) e^{-\Phi \Phid/3}+W(\Phi)\Big]+h.c. \nn \\
&& -\frac{1}{8} \int \d^2\Theta 2\mac{E} (\bcD^2-8R) \cD \Phi \cD \Phi \bcD \Phid \bcD \Phid \, T_{\rm{DBI}} +h.c.,  \label{cern1}
\eea
where we have added a superpotential $W$ and let
\begin{eqnarray}
&&16T_{\rm{DBI}} = \\
&& \frac{f(\Phi,\Phid) }{1 + f \pt \Phi \cdot \pt \Phid e^{K/3} + \sqrt{(1 + f \pt \Phi \cdot \pt \Phid e^{K/3})^2 - f^2  (\pt \Phi)^2 (\pt \Phid)^2 e^{2K /3}}}. \nonumber
\end{eqnarray}
Here, factors of $e^{K/3}=e^{\Phi\Phid /3}$ have been introduced so as to compensate for the Weyl rescaling that must be performed to go to Einstein frame. In components fields, action \eqref{cern1} becomes
\bea
\frac1e {\cal L}_{\rm{DBI}} &=& -\frac12 \cR- \frac{1}{f} \left( \sqrt{{\det}(g_{mn} + f\, \pt_m A \pt_n A^*)} -1 \right) \nn \\ 
&& +  e^{AA^*/3} |F|^2 + \frac{}{}e^{2AA^*/3}[F (D_A W) + F^{*} (D_A W)^*]  + 3 e^{AA^*} |W|^2 \nn \\ &&  -32 \frac{}{}e^{AA^*/3} |F|^2 |\pt A|^2 \, {\cal{T}}_{\rm{DBI}} \nn \\ && + 16 e^{2AA^*/3} |F|^4 \, {\cal{T}}_{\rm{DBI}} \ .
\eea
The auxiliary field $F$ now obeys the equation of motion
\be
F+ e^{AA^*/3} (D_A W)^* + 32 F \, \mac{T}_{\rm{DBI}} (e^{AA^*/3} |F|^2 - |\pt A|^2) = 0 \ . \label{EqofMotionFDBI}
\ee
We will consider three regimes of interest here, leaving a more detailed study to future work. When $f$ is small, that is, when the higher-derivative terms are subdominant, one can apply the results of Subsection \ref{sectionsmallT}.  Then the ordinary branch solution for $F$ leads to small correction terms, the resulting Lagrangian being
\bea
&& \frac1e \mac{L}_{\rm{DBI,ordinary,\mac{T}\rightarrow 0}} \nn \\ &=& -\frac12 \cR - \frac{1}{f} \left( \sqrt{{\det}(g_{mn} + f\, \pt_m A \pt_n A^*)} -1 \right)  \nn \\ && -  \frac{2f e^{K} |D_A W|^2}{1 + f\, |\pt A|^2 + \sqrt{(1 + f\, |\pt A|^2)^2 - f^2 \, (\pt A)^2 (\pt A^*)^2}}  |\pt A|^2 \nn \\ \\&& -\frac{}{}e^K (|D_A W|^2 - 3 |W|^2)  \nn \\ && 
  +  \frac{f e^{2K} |D_A W|^4}{1 + f\, |\pt A|^2 + \sqrt{(1 + f\, |\pt A|^2)^2 - f^2 \, (\pt A)^2 (\pt A^*)^2}}
\eea
with $K=AA^*.$ Both the kinetic and the potential terms receive corrections, which are, however, necessarily small in the limit under consideration. Nevertheless, it will be important to include such terms when working out the detailed predictions of phenomenological or cosmological models based on DBI actions.  

From the point of view of the present paper, as well as for applications to models of DBI inflation, the regime where $f$ is large is the most interesting one. Indeed, a special feature of the DBI action is that for large $f$-- and restricting to fields that depend only on time --the scalars get slowed down and obey a stringent upper speed limit. By inspection of the Lagrangian, one can see that this upper limit corresponds to 
\be
f \, |\dot{A}|^2  \leq \frac12.
\ee
The ``relativistic'' limit, where this bound is (approximately) saturated, is clearly of particular importance to models of DBI inflation, as it can ensure slow-roll even in relatively steep potentials. However, precisely because the kinetic term becomes small as $f$ becomes large, the relativistic limit does not immediately correspond to the large $\mac{T}$ limit of Subsection \ref{sectionlargeT}. Indeed, $\mac{T}_{DBI}$ becomes large, but the higher-derivative terms nevertheless do not become completely dominant. For this reason, one cannot blindly apply the formulas of Subsection \ref{sectionlargeT}. Instead, one must start again from the equation of motion (\ref{FFFEquation}) for the auxiliary field $F$. For simplicity, we restrict our analysis to a single real scalar $\phi = \sqrt{2} \rm{Re}(A).$ Then $K = \phi^2/2.$ In the present context, the auxiliary field equation of motion reduces to
\be
F^3 + \frac{3}{4f}e^{-K/3} \frac{(D_A W)^*}{D_A W}F + \frac{1}{4f} \frac{(D_A W)^{*2}}{D_A W} =0, \label{FFFEq2}
\ee
where we have used
\be
16 \mac{T}_{\rm{DBI}} = \frac{f}{1 + \frac12 f\, (\pt \phi)^2 + \sqrt{1 + f\, (\pt\phi)^2}} \approx 2 f 
\ee
for the relativistic limit $f\,(\pt\phi)^2 \approx -1$. In the limit that $(f e^{K} |D_A W|^2)^{1/3}$ is large, the solution to Eq. (\ref{FFFEq2}) is given by
\be
F \approx - \left( \frac{(D_A W)^{*2}}{4f \, D_A W} \right)^{1/3}.
\ee
Substituting this solution back into the Lagrangian gives
\bea
\frac1e \mac{L}_{\rm{DBI,relativistic}} &=& -\frac12 \cR - \frac{1}{f} \left( \sqrt{1+ f\, (\pt\phi)^2} -1 \right)  \nn \\ && -\frac32 \frac{e^{K} |D_A W|^2}{[4f\, e^{K} |D_A W|^2]^{1/3}} + 3 e^{K} |W|^2, 
\eea
where the higher corrections are of order ${\cal O}((f\, e^{K} |D_A W|^2)^{-2/3}).$ Remarkably, the first part of the potential is subleading, and the dominant contribution to the potential, namely $-3 e^{K} |W|^2,$ is {\it negative}! Thus, for this simple, single-field supergravity realization of the DBI action, inflation cannot occur in the relativistic regime. As the higher-derivative terms become increasingly important, the potential becomes correspondingly more negative. A question, which we leave to future work, is whether this limitation can be overcome by considering either more fields or different supergravity extensions of the DBI model\footnote{In \cite{Sasaki:2012ka} a study of supersymmetric DBI inflation was undertaken, where the authors also highlighted the importance of the cubic equation of motion for $F$, and where they considered similar limits to those considered here. It was claimed that in the small $f$ limit (large $T$ in their notation) inflation cannot occur, but that in the large $f$ limit, with very small $\dot\phi^2,$ it could. They also excluded relativistic DBI inflation, but for reasons different than ours. Our results differ rather significantly, which can in part be traced back to the fact that we are performing the analysis in supergravity, whereas the authors of \cite{Sasaki:2012ka} considered a hybrid approach where the formulae of global supersymmetry were simply added to an Einstein-Hilbert term.}.

Finally, an interesting theory can arise in the absence of a superpotential. In that case, as discussed in detail in Subsection \ref{sectionnoW}, apart from the trivial solution $F_0=0$-- which leads to the standard DBI theory --there exists a new solution satisfying
\be
|F_{\rm{new}}|^2 = e^{-AA^*/3}\Big(\frac12 |\pt A|^2 - \frac{1}{2f}(1+\sqrt{(1+ f|\pt A|^2)^2 - f^2  (\pt A)^2 (\pt A^*)^2}\Big) \label{FDBInew} 
\ee
When substituted into the action, we obtain the Lagrangian 
\bea
\frac1e {\cal L}_{\rm{DBI,new}} &=& -\frac12 \cR - \frac{1}{4} |\pt A|^2 \nn \\ &&  - \frac{1}{4f} \left( \sqrt{{\det}(g_{mn} + f\, \pt_m A \pt_n A^*)} -1 \right)  -\frac{1}{2f}  \\ && + \frac{f\big((\pt A)^2 (\pt A^*)^2 - |\pt A|^4\big)}{1 + f\, |\pt A|^2 + \sqrt{(1 + f\, |\pt A|^2)^2 - f^2 \, (\pt A)^2 (\pt A^*)^2}}. \nn
\eea
Although the ordinary kinetic term has disappeared, a new kinetic term, as well as a new potential, have regrown via the higher-derivative interactions! In the purely time-dependent case, where $A = (\phi(t) + \xi(t))/\sqrt{2}$, the theory reduces to
\bea
\frac1e {\cal L}_{\rm{DBI,new}} &=& -\frac12 \cR + \frac{1}{4} X - \frac{1}{4f} \left( \sqrt{1-2fX} -1 \right)  -\frac{1}{2f}, \label{DBInewt}
\eea
where $X = \frac12 (\dot\phi^2 + \dot\xi^2).$ For the type IIB string theory case of a $D3$-brane moving in a warped throat Calabi-Yau geometry, one has $f \propto (\phi^2 + \xi^2)^{-2}.$ Hence, in this setting our new theory-- which has both an ordinary kinetic term and an additional DBI term --would contain a non-vanishing potential proportional to $(\phi^2 + \xi^2)^2$. Thus, even in the absence of a superpotential, an effective potential is generated. One must remember, 
however,  that for consistency the right-hand side of (\ref{FDBInew}) must be positive. It is straightforward to convince oneself that for scalar fields which depend only on time this requires that we take $f<0$ and, hence, the potential is required to be negative. If we allow the fields to depend on space as well, then the right-hand side of (\ref{FDBInew}) can be positive when the fields develop large spatial gradients, even when $f$ is positive. Either way, however, this new branch of the theory does not allow for a phase of inflation to occur. An interesting question is what prevents the theory from dynamically reaching the ``forbidden'' field values, where $|F|^2$ would become negative. We leave this open question for future work.

\vspace{.5cm}
\noindent{\underline{Summary}:} {\it Our formalism allows one to construct a supergravity version of the DBI action. When the higher-derivative terms are small, we obtain correspondingly small corrections to the DBI Lagrangian and to the potential. In the most interesting case, where the higher-derivative terms significantly influence the dynamics, we find that the potential again becomes
\be
V_{\mac{T}\rightarrow \infty}=-3 e^K |W|^2 \ ,
\ee
which is everywhere negative.
This result represents a serious challenge to models of DBI inflation where the relativistic regime of the theory is exploited.
In the absence of a superpotential, the new branch of the supergravity DBI theory generates a potential, but, curiously, this theory either requires the potential to be negative (without restricting the types of solutions that the scalars can admit) or, if the potential is positive, it requires the scalars to develop large spatial gradients.
}

\section{Conclusions and Outlook} \label{sectionconclusion}

In this paper, we presented a formalism that allows one to obtain an $\mac{N}=1$ supergravity extension of any scalar field theory with higher-derivative kinetic terms. This was accomplished by constructing a superfield-- quartic in chiral scalars --which contains the term $(\pt\phi)^4$ and, when the fermions are set to zero, consists entirely of its top component. Thus, when multiplied by any other superfield, the resulting Lagrangian contains only the lowest component of the multiplicative factor. This property enables one to directly construct a supergravity extension any higher-derivative scalar field term of interest. Moreover, as discussed in the Appendix, our supergravity extension of $(\pt\phi)^4$ is likely to be the unique one that does not modify the gravitational sector of the theory-- thus rendering our construction particularly pertinent. For this reason, studying the properties of the auxiliary fields in this context, which are crucial to the structure of supergravity, is important. This was carried out, in detail, in this paper.

In our formalism, despite the inclusion of an arbitrarily high number of spacetime derivatives, the auxiliary fields do not have kinetic terms and, therefore, continue to satisfy algebraic equations of motion. We point out that this is a highly non-trivial property, which renders the treatment of the auxiliary fields straightforward. Be this as it may, there is one new, and important, property of our formalism. That is, although the auxiliary fields $F$ satisfy an algebraic equation of motion, that equation is now cubic-- as opposed to the linear equation in the usual second order kinetic theory.  Hence, this equation admits up to three distinct solutions. We have shown that these solutions lead to different theories that cannot dynamically transition from one to another. One solution is directly related to the one ordinarily obtained in the absence of higher-derivative terms. This leads to corrections to both the kinetic and potential terms when substituted into the action. We have examined these corrections in different limits.  When the higher-derivative terms are small, the corrections are correspondingly small, but need to be taken into account when making precise predictions in phenomenology and cosmology. In the limit that the higher-derivative terms become large, the effect of eliminating the auxiliary field is to suppress certain contributions to the potential. The result is that the negative term $-3e^K|W|^2$ becomes the dominant contribution to the potential energy. Thus, in the large higher-derivative limit, supergravity manifests once more its predilection for negative potentials. This feature implies that the supergravity implementation of inflationary and $k$-essence models-- such as DBI inflation --that rely on higher-derivative kinetic terms in an essential way  become more challenging. 

In addition to this ``usual'' solution for $F$, there exist up to two new solutions. These lead to theories with very unusual properties, which we have only started exploring in the present paper. For example, these new branches seem to prefer solutions with substantial spatial gradients in the scalar fields, and can lead to positive potentials. Moreover they can do this even in the absence of a superpotential. These curious theories, whose physical relevance is not clear yet, form an interesting topic for further research.

This work has many foreseeable applications. Most importantly, we hope that our results can be used to bridge the gap between standard model building in cosmology and full-blown string compactifications, leading to well-motivated effective theories of early universe dynamics. In this context, it will be interesting to investigate in more detail models of DBI inflation and $k$-inflation, as well as other models of brane dynamics such as the Galileons and their extensions. Furthermore, it will be enlightening to find out whether null energy violating models, such as the ghost condensate, can be realized in a supergravity context. We hope to explore these topics in the near future.

\section*{Acknowledgments}

We would like to thank Ilarion Melnikov for useful discussions. M.K. and J.L.L. are thankful to the University of Pennsylvania for its warm hospitality during the course of this work. M.K. and J.L.L. also gratefully acknowledge the support of the European Research Council via the Starting Grant numbered 256994. B.A.O. is supported in part by the DOE under contract No. DE-AC02-76-ER-03071 and the NSF under grant No. 1001296.

\appendix

\section{Relationship to the Work of Baumann and Green}

An initial study of an effective supergravity theory of higher-derivative scalar fields was performed by D. Baumann and D. Green in \cite{Baumann:2011nm} (other earlier works of interest include \cite{Buchbinder:1988yu,Buchbinder:1994iw,Banin:2006db,Brandt:1993vd,Brandt:1996au,Antoniadis:2007xc}), and applied to certain cosmological questions in \cite{Baumann:2011nk}. These authors based their formalism on a different supergravity extension of $(\pt\phi)^4,$ given by
\be
\mac{L}_{\rm{BG}} = -\frac{1}{32} \int \d^2\Theta 2\mac{E} (\bcD^2-8R) (\Phi - \Phid)^2 \cD_a \Phi \cD^a \Phid. \label{BG}
\ee
The component expansion of this Lagrangian contains $(\pt\phi)^4,$ as desired. It also contains terms that are of a rather different character than those considered in our work. For example, the above superfield generates derivative couplings to the Ricci tensor of the form
\be
\xi^2 (\pt\phi)^2 \cR, \qquad \xi^2 \cR^{mn} \pt_m\phi \pt_n\phi.
\ee
Such couplings modify the gravitational part of the theory in a non-trivial manner. This is both interesting for phenomenology and difficult for calculations, as one cannot Weyl rescale such terms away. For this reason, it becomes more difficult to interpret the resulting theory. The component expansion of (\ref{BG}) also contains a term
\be
\xi^2 |\pt F|^2 \ ,
\ee
which makes the ``auxiliary'' field become propagating. This implies that the field $F$ cannot be eliminated as usual, but must be retained as a dynamical, propagating degree of freedom. For these reasons, the term (\ref{BG}) takes us outside the class of theories we want to consider in the present work. However, in a general supersymmetric effective field theory, such a term could also be present. Consequently a study of its properties and phenomenological consequences is certainly of interest.

A final remark. In \cite{Khoury:2010gb} arguments were given that, in global supersymmetry, the two superfield expressions $D\Phi D \Phi \bar{D} \Phid \bar{D} \Phid$ and $(\Phi - \Phid)^2 D_a \Phi D^a \Phid$ are the only ``clean'' supersymmetric extensions of $(\pt\phi)^4.$ By this we mean that they generate $(\pt\phi)^4$, but no additional terms containing only $\phi.$ We now see that in supergravity these two superfield expressions differ in their coupling to gravity, with $\cD\Phi \cD \Phi \bcD \Phid \bcD \Phid$ leading to minimal coupling while $(\Phi - \Phid)^2 \cD_a \Phi \cD^a \Phid$ gives additional derivative couplings.

\section{Comment on K\"{a}hler Invariance}

In the usual theory of chiral superfields coupled to supergravity, invariance under K\"{a}hler transformations plays an important role. Thus, one may wonder if this symmetry also restricts higher-derivative terms. Since the same question arises in chiral models with global supersymmetry, we will analyze the question in that simpler context. Super-K\"{a}hler transformations correspond to a shift of the vector K\"{a}hler superfield
\be
K(\Phi^{i},{\Phi}^{\dagger i*}) \rightarrow K(\Phi^{i},\Phi^{\dagger i*}) + C(\Phi^{i}) + C^*(\Phi^{\dagger i*}) \ ,
\label{dude1}
\ee
where $C$ is an arbitrary holomorphic function of chiral superfields and $C^*$ is its conjugate. Since the usual two-derivative chiral superfield Lagrangian is 
\be
{\cal L} = \int \d^2 \th \d^2 \bar\th K \ ,
\ee
invariance under super-K\"{a}hler transformations is almost a trivial statement, following from the fact that the top component of any chiral superfield is a total spacetime derivative-- given by $\frac14 \Box$ of its lowest component. 

Less trivial, however, is the following. Note that the super-shift \eqref{dude1} induces the scalar K\"ahler transformation
\be
K(A^{i},A^{i*}) \rightarrow K(A^{i},A^{i*}) + C(A^{i}) + C^*(A^{i*}) \label{K2}
\ee
in the lowest component of $K(\Phi^{i},{\Phi}^{\dagger i*})$. Furthermore, the $\th^{2}{\bar\th}^{2}$ component of $K(\Phi^{i},\Phi^{\dagger i*})$,
which gives the two-derivative component field Lagrangian, contains the non-linear sigma model $K_{,A^{i}A^{j*}}{\cal{D}}^{m}A^{i}{\cal{D}}_{m}A^{j*}$ for the scalar fields $A^{i}$. Under the K\"ahler transformation \eqref{K2}, this is invariant since $K$ appears with mixed second derivatives. This invariance-- unlike the total divergence terms --is very non-trivial and corresponds geometrically to the target space of the scalar fields being a complex K\"ahler manifold with K\"ahler metric $g_{ij*}=K_{,A^{i}A^{j*}}$.

Now consider higher-derivative contributions to the Lagrangian. As discussed in the text, in flat superspace these take the form 

\begin{equation}
{\cal{L}}_{\rm h-d}= \int \d^2 \th \d^2 \bar\th D \Phi^i D \Phi^j \bar{D} \Phi^{\dagger k*} \bar{D} \Phi^{\dagger l*}T_{ijk*l*} \ .
\label{dude2}
\end{equation}
To maintain sigma-model diffeomorphism invariance, it is necessary that $T_{ijk*l*}$ transform as a tensor on the complex scalar manifold. Furthermore, consistency with the K\"ahler manifold required by the two-derivative Lagrangian implies that this tensor be chosen invariant under K\"ahler transformations \eqref{dude1}. An example of this is to take $T_{ijk*l*}\propto (K_{,\Phi^{i}\Phi^{\dagger j*}})^{2}$ times a K\"ahler invariant scalar superfield--as was done in the text for the case where all fermions are set to zero.
Thus, the requirement that the action be K\"ahler invariant {\it does} restrict the higher-derivative terms. Finally, these arguments carry over directly to curved superspace and, hence, higher-derivative chiral superfield Lagrangians coupled to ${\cal{N}}=1$ supergravity.

\bibliographystyle{unsrtabbrv}
\bibliography{master}
\end{document}